\documentclass[article,fleqn,showpacs,nofootinbib]{revtex4}
\usepackage{graphicx}
\usepackage{amssymb}
\usepackage{amsmath}
\usepackage{bm}
\usepackage{color}

\newcommand{\beq}{\begin{equation}}
\newcommand{\eeq}{\end{equation}}
\newcommand{\beqa}{\begin{eqnarray}}
\newcommand{\eeqa}{\end{eqnarray}}
\newcommand{\bseq}{\begin{subequations}}
\newcommand{\eseq}{\end{subequations}}

\newcommand{\mn}{{\mu \nu}}

\newcommand{\cE}{\mathcal{E}}

\begin{document}

\begin{flushright}
{\tt arXiv:YYMM.NNNN}
\end{flushright}

\setcounter{page}{1}
\title{Vulcanized Vortex}
\author{Inyong \surname{Cho}}
\email{iycho@skku.edu}
\affiliation{Department of Physics and BK21 Physics Research Division,
Sungkyunkwan University, Suwon 440-746, Korea}
\author{Youngone \surname{Lee}}
\email{youngone@daejin.ac.kr}
\affiliation{Department of Physics,
Daejin University, Pocheon, Gyeonggi 487-711, Korea}

\date[]{}

\begin{abstract}
We investigate vortex configurations
with the ``vulcanization" term inspired by
the renormalization of $\phi_\star^4$ theory
in the canonical $\theta$-deformed noncommutativity.
We focus on the classical limit of the theory
described by a single parameter
which is the ratio of the vulcanization and
the noncommutativity parameters.
We perform numerical calculations and
find that nontopological vortex solutions exist
as well as Q-ball type solutions,
but topological vortex solutions are not admitted.
\end{abstract}



\maketitle

\section{Introduction}
In the quantum theory of spacetime,
we have noncommutative algebras of observables
which correspond to commutative algebras in classical theory.
Although we do not know the exact noncommutative algebra,
or the commutation relations between related observables yet,
studying some models of noncommutative spacetime
 would be instructive in that
it gives some hints for true quantum spacetime physics.

Among many proposed noncommutative models,
the canonical $\theta$-deformed noncommutativity~\cite{Doplicher:1994tu}
and the $\kappa$-deformed noncommutativity~\cite{Lukierski:1991pn}
have been widely studied.
Our work in this paper is based on
the canonical noncommutative spacetime.
The commutation relation between coordinates is given by
\begin{equation}
\label{nc1}
\left[x^\mu, x^\nu\right] = i \theta^{\mn},
\end{equation}
where $\theta^{\mu\nu}$ is a constant antisymmetric matrix.
By a suitable transformation, we can always write $\theta^{\mu\nu}$
in the form of
\begin{equation}
[\theta^{\mu\nu}] =
\left(
\begin{array}{cccc}
  0 & \theta_E & 0 & 0 \\
  -\theta_E & 0 & 0 & 0 \\
  0 & 0 & 0 & \theta_B \\
  0 & 0 & -\theta_B & 0
\end{array}
\right),
\label{thetamat}
\end{equation}
which exhibits SO(2)$\times$SO(2) structure.

Quantum field theory in this canonical noncommutative spacetime (NCQFT)
has many characteristic aspects
when compared to commutative theory.
Among those aspects,
the most significant pathology is
the nonrenormalizability of the theory.
Minwalla et.al. pointed out
the mixing of different scales of the theory known as
UV/IR mixing in the canonical NCQFT~\cite{Minwalla:1999px}.
The perturbative analysis of the canonical NCQFT shows
infinitely many infrared divergencies which cannot be reabsorbed
to the parameters of the original theory.

This problem of nonrenormalizability has been recently resolved by
Grosse and Wulkenhaar~\cite{Grosse:2004yu}.
By adding the so-called ``vulcanization" term to the action
they could successfully show the renormalizability of the theory.
The vulcanized action of the $\phi_\star^4$ theory has the form,
\begin{equation}
\label{gwaction}
S[\phi] = \int d^4x \sqrt{g} \left[
{1\over 2} \partial_\mu\phi \star \partial^\mu\phi
+{\Omega^2 \over 2} (\tilde{x}_\mu\phi)\star (\tilde{x}^\mu\phi)
+{m^2\over 2}\phi\star\phi
+{\lambda\over 4} \phi\star\phi\star\phi\star\phi \right],
\end{equation}
where the second term is the vulcanization term
with $\tilde{x}^\mu \equiv 2 (\theta^{-1})^{\mu\nu}x_\nu$,
and the metric is Euclidean.
Their conclusion was that
one should change the free propagator
to get a renormalizable NCQFT.
The guideline (using Langmann-Szabo symmetry~\cite{Langmann:2002cc}) was
to incorporate the mixing of ultraviolet physics
with infrared physics in each order of the perturbation.
It was later proved in Ref.~\cite{Gurau:2005gd} that
the value of the coupling constant $\Omega$ is not restricted.
It simply needs to be any value which is nonzero positive.

If we require renormalizability as a guideline,
additional terms in the noncommutative field theory action
will be inevitable as we have seen in the vulcanized theory.
In the classical limit ($\theta_{i=E,B}\rightarrow 0$),
the action \eqref{gwaction} takes the form,
\begin{equation}
\label{claction}
S[\phi] = \int d^4x \sqrt{g} \left[
{1\over 2} \partial_\mu\phi  \partial^\mu\phi
+{\Omega^2 \over 2} \tilde{x}_\mu\tilde{x}^\mu\phi \phi
+{m^2\over 2}\phi^2
+{\lambda\over 4} \phi^4 \right].
\end{equation}
The noncommutativity parameter $\theta_i$ is hidden in the vulcanization term,
\begin{equation}
{\Omega^2 \over 2} \tilde{x}_\mu\tilde{x}^\mu\phi \phi
=2{\Omega^2 } \left[ {(x_E)^a (x_E)_a \over \theta_E^2}
+ {(x_B)^a (x_B)_a \over \theta_B^2} \right]\phi\phi,
\label{VulEB}
\end{equation}
where $(x_i)^a$ is the coordinate corresponding to the electric/magnetic components
in Eq.~\eqref{thetamat}.
When $\theta_i =0$, the corresponding term in Eq.~\eqref{VulEB} is absent.
The theory in the classical limit is now described by the  parameter
$\Omega_{\theta_i} = \Omega/\theta_i$,
which is the ratio of the vulcanization and the noncommutativity parameters.
Since $\Omega$ can be any positive value,
we can always make $\Omega_{\theta_i}$ be finite for any small $\theta_i$.
In this picture, the classical limit of the vulcanization theory
is never the same with the ordinary classical action,
so it is very interesting to study this limit.
To understand this classical limit,
it will be helpful to investigate classical objects
in the presence of the vulcanization term.

Since the vulcanization term is coordinate-dependent,
it is suggestive to consider an inhomogeneous field configuration
resulting from the theory.
Interesting candidates for such a configuration would be
(non)topological solitons.
In this article, we shall introduce a complex scalar field
which is associated with global U(1) symmetry.
The possible solitonic candidates are topological~\cite{VS}
or nontopological~\cite{Kim:1992mm} vortices,
and Q-ball type nontopological solitons~\cite{Friedberg:1976me,TDLee}.
Topological solitons arise in the field theory of symmetry breaking.
Their outer boundary takes the vacuum-expectation value
in the broken-symmetry state.
The vacuum is specified by a nontrivial homotopy group.
This type of soliton is topologically stable since
there is no continuous transformation that deforms them homotopically
to a trivial solution.
Nontopological solitons, on the other hand, can arise with the theory of unbroken symmetry.
The outer boundary acquires the value of the vacuum which is not
nontrivial homotopically.
Their stability is provided by a conserved U(1) charge.
They are usually the lowest-energy configuration
of the theory.

We perform numerical calculations in order to see
if there exist field configurations which satisfy
boundary conditions for the various types of solutions above.
In our model of the vulcanized theory,
we find that there may exist nontopological vortices
and Q-ball type solitons of which integrated energy is finite.
However, the vortex solution which meets the
topological boundary condition does not exist.
In Sec.~II, we present the setup of the vulcanization model
with global U(1) symmetry.
In Sec.~III, we numerically search various (non)topological
solutions for the field and analyze them.
In Sec.~IV, we conclude.

\section{Setup}
In this section, we present a vortex model
motivated by the vulcanized noncommutative theory.
In order to consider a vortex configuration,
we introduce a complex scalar field.
Then the classical action~\eqref{claction} becomes
\begin{equation}
S[\phi] = \int d^4x \sqrt{-g} \left[
-{1\over 2} \partial_\mu \bar\phi \partial^\mu\phi
-{\Omega^2 \over 2} \tilde{x}_\mu\tilde{x}^\mu \bar\phi\phi
-{m^2\over 2}\bar\phi\phi
-{\lambda\over 4} (\bar\phi\phi)^2\right],
\label{Sphi}
\end{equation}
where the metric is now Lorentzian.
We do not face any trouble which arises in the Wick rotation
from Euclidean to Lorentzian
since we shall consider only spatial noncommutativity in what follows.
We keep the SO(2) structure only in the spatial part on the $(x,y)$-plane,
i.e., $\theta_E=0$ and $\theta_B=\theta$.
Then there is a translational symmetry in the $t$- and $z$-directions,
and a circular symmetry in the $(x,y)$-plane.
This corresponds to the cylindrical symmetry in 3D space.

In the cylindrical coordinate system $\{t,z,\rho,\vartheta\}$,
the field equation then becomes
\begin{equation}
\nabla_\mu\partial^\mu\phi =
\left[ -{\partial^2 \over \partial t^2}
+{1\over \rho}{\partial \over \partial\rho}\left(\rho {\partial\over \partial\rho}\right)
+{1\over \rho^2}{\partial^2\over \partial\vartheta^2}
+{\partial^2\over\partial z^2} \right]\phi
=\lambda\phi\bar\phi\phi +m^2\phi
+{4\Omega_\theta^2} \rho^2\phi,
\label{phieq2}
\end{equation}
where $\Omega_\theta = \Omega/\theta$.

For the vortex configuration, we shall consider both topological and nontopological
solutions. The ansatz for such a field can be
\begin{equation}
\phi (t,\rho,\vartheta) = e^{i\omega t} e^{in\vartheta} f(\rho),
\label{phi}
\end{equation}
where $f(\rho)$ is the radial profile of the field,
$n$ is an integer representing vorticity,
and $\omega$ is a number which is associated with the conserved U(1) charge.
Since the model possesses global U(1) symmetry,
there is a conserved Noether current
$j^\mu = (i/2) [\bar\phi\partial^\mu\phi -(\partial^\mu\bar\phi)\phi ]$
whose charge is
$Q=\int d^2x j^0=\int d^2x \omega |\phi|^2$.
While the stability of topological vortices is provided by the topological reason,
i.e., nontrivial structure of the vacuum,
that of the nontopological ones is provided by the conserved U(1) charge
related with $\omega$.

Plugging the ansatz~\eqref{phi} in the field Eq.~\eqref{phieq2},
the field equation for the radial profile becomes
\begin{equation}
\left( {d^2 \over d\rho^2} + {1\over\rho}{d\over d\rho}- {n^2\over \rho^2}\right) f
= \lambda f^3 - (\omega^2- m^2)f +{4\Omega_\theta^2 }\rho^2 f.
\label{feq1}
\end{equation}
The vulcanization term is interpreted as a kind of source term,
and the effective potential which constrains the field is given by
\begin{equation}
V_{\rm eff} = V - {\omega^2 \over 2} f^2,
\quad \mbox{ where } V = {\lambda\over 4}f^4 + {m^2 \over 2} f^2 +V_0.
\label{V}
\end{equation}
The constant $V_0$ is absent in the original action~\eqref{Sphi},
but its introduction does not change the picture of the model.
The value of $V_0$ simply shifts the magnitude of the vacuum energy
which is indeed required to be zero for nontopological vortices.

\begin{figure}[t]
\includegraphics[width=60mm]{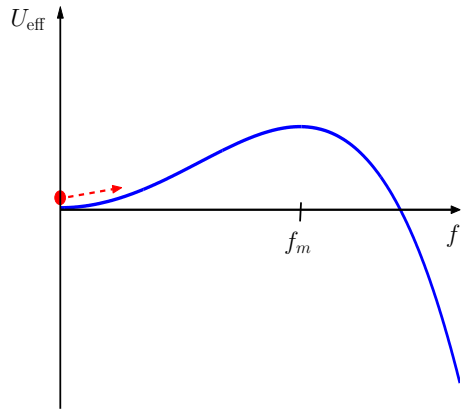}
\caption{Plot of the effective potential.
The local maximum is at $f_m=\sqrt{(\omega^2-m^2)/\lambda}$.
For nontopological vortices, the field starting from zero with a finite velocity
should come back to the original point with zero velocity and acceleration.}
\label{fig1}
\end{figure}

\section{Soliton solutions}
In this section, we search soliton solutions of the field Eq.~\eqref{feq1}.
In order to examine the existence of solutions for given boundary conditions,
we treat the field equation as a dynamical one
of which the solution is interpreted as the motion of the
particle subject to the Newtonian equation.
Then the field profile $f$ means the position of the particle, and
the radial coordinate $\rho$ plays the role of time.
The field Eq.~\eqref{feq1} can be recast into the form
\begin{equation}
{d^2f \over d\rho^2} = - {1\over\rho}{df\over d\rho}+ {n^2\over \rho^2} f
+ \lambda f^3 - (\omega^2- m^2)f +{4\Omega_\theta^2}\rho^2 f,
\label{feq2}
\end{equation}
where the left-hand side is acceleration
and the terms on the right-hand side play a role of forces.
The first term on the right-hand side is a friction
proportional inversely to time,
and the second term is a time-dependent repulsion which decays in time.
The combination of the third and the fourth term is the conservative force subject to
the effective potential $U_{\rm eff} = -V_{\rm eff}$.
The last vulcanization term plays the same role with the
second term (repulsive-force) but amplifies in time.

For right solutions, the motion subject to the effective potential
$U_{\rm eff}$ shown in Fig.~1 (we need $\omega^2 >m^2$)
and to the forces in Eq.~\eqref{feq2}, should meet
the appropriate boundary conditions for soliton configurations.
For soliton configurations, we consider
topological vortices, nontopological vortices, and Q-ball type solitons
which have different types of boundary conditions.
We shall deal with them separately.

\subsection{Topological vortex}
The formation of topological vortices is
associated with symmetry breaking.
It is stable due to the topological reason
and the frequency term is not required, $\omega =0$.
The corresponding potential is a Mexican-hat type
given by the tachyonic mass, $m^2<0$.
At the center of the vortex,
the field stays at the symmetry state ($f=0$),
and approaches the broken-symmetry state ($f = |m|/\sqrt{\lambda}$) asymptotically.

For the topological vortex solutions with the boundary conditions above,
the field starts from $f (\rho=0) =0$ with a finite velocity and should reach
the top of the effective potential in Fig.~1, $f(\rho\to \infty)\to |m|/\sqrt{\lambda}$.
At the top of $U_{\rm eff}$, the acceleration and velocity should vanish.
It is not difficult to see from Eq.~\eqref{feq2}
that this boundary condition cannot be met.
With finite $f$ and velocity $df/d\rho$, as $\rho$ becomes large,
the force terms approach zero except the vulcanization term which diverges.
This makes the acceleration diverge, and the motion fails to meet the
topological boundary condition.

More analytically, for vanishing acceleration ($d^2f/d\rho^2 =0$)
at the top of the potential ($\partial U_{\rm eff}/\partial f =0$),
the field Eq.~\eqref{feq2} and the solution to it become
\begin{equation}
- {1\over\rho}{df\over d\rho}+ {n^2\over \rho^2} f
+{4\Omega_\theta^2}\rho^2 f = 0
\quad\Rightarrow\quad
f=c_1 \rho^{n^2} \exp\left({\Omega_\theta^2 } \rho^4\right).
\end{equation}
It is easy to see that the solution is an increasing function
of $\rho$ and cannot reach a nonzero finite value asymptotically.
Locally at a finite $\rho$ one can have $f(\rho_*) = |m|/\sqrt{\lambda}$,
but its velocity is nonzero there.
As a result, the vulcanized noncommutative model does not admit
a topological vortex solution.

\subsection{Nontopological vortex}
Nontopological solitons arise in the unbroken-symmetry theory~\cite{Friedberg:1976me,TDLee},
and their stability is provided by the conserved U(1) Noether charge.
The nontopological vortex, in particular, is an object
produced in such a theory with vorticity~\cite{Kim:1992mm}.
For the nontopological vortex, therefore,
both $n$ and $\omega$ are turned on in the field ansatz~\eqref{phi},
which indicates that there is a constant angular momentum.
The boundary conditions for such a vortex are
$f(\rho=0) = 0$ and $f(\rho\to\infty) \to 0$;
the field resides in the symmetry state in both boundaries.
The shape of the field in the intermediate region between boundaries
represents ringlike matter rotating about the center with an angular momentum.

Dynamically, the particle subject to the effective potential $U_{\rm eff}$
starts from $f(\rho=0) = 0$ with some velocity and climbs up the hill.
Then it has to turn back and to settle down at the original position $f=0$
with zero acceleration and velocity at $\rho\to\infty$.
From Eq.~\eqref{feq2}, as $\rho\to\infty$,
the friction and the repulsion decay to zero.
As $f\to 0$, the conservative force by $U_{\rm eff}$ becomes zero.
The remaining vulcanization term $\propto \rho^2f$ can also approach zero
if $f$ decays faster than $1/\rho^2$.
The vulcanization term possibly provides the zero-acceleration boundary condition.
However, the behavior of the velocity $df/d\rho$ remains undetermined.
If we assume that $f$ decays faster than $1/\rho^2$,
the $f^3$-term in Eq.~\eqref{feq2} becomes most negligible.
Then, there exists an approximate solution at large $\rho$,
\begin{equation}
f(\rho) \approx {c_1\over \rho}
{\rm WhittakerW} \left( {\omega^2-m^2 \over 8\Omega_\theta},{n\over2},
{2\Omega_\theta}\rho^2 \right),
\label{Whittaker}
\end{equation}
which decreases to zero rapidly as well as its velocity and acceleration.
Therefore, it is very probable that there exists a vortex solution with
nontopological boundary conditions.
What is remaining is to check if this approximate large-$\rho$ solution
is obtained when the field equation is integrated from the core region
with appropriate initial conditions.
In order to see this, we need to perform numerical calculations.

For numerical calculations, we search the solution which approaches $f=0$
asymptotically at large $\rho$ with shooting at small $\rho$ by tuning the velocity.
In the vicinity of $\rho =0$, the field can be expanded as
\begin{equation}
f(\rho) = f_1 \rho^n \left\{
1 - {\omega^2-m^2 \over 4(n+1)}\rho^2
+ {1\over 8(n+2)} \left[
{(\omega^2-m^2)^2 \over 4(n+1)} + {4\Omega_\theta^2}
+\lambda f_1^{2}\delta^n_1 \right] \rho^4
+{\cal O}(\rho^6) \right\}.
\label{IC}
\end{equation}
By varying $f_1$ we search the solution which satisfies the
boundary conditions, $f\to 0$ and $df/d\rho\to 0$ at large $\rho$.

In order to satisfy the boundary conditions at both ends,
the field $f$ should increase first and then decrease.
Before we perform numerical calculations,
let us discuss necessary conditions for the turning of the field
to have such a configuration.
At the moment of turning, $\rho =\rho_t$, the velocity becomes zero, $(df/d\rho)(\rho_t) =0$.
The acceleration at this moment should be negative in order for the field
to come back to the origin,
\begin{equation}
\left. {d^2f \over d\rho^2} \right|_{\rho=\rho_t}
= \left[ {n^2\over \rho^2} f + \lambda f^3 -
(\omega^2- m^2)f +{4\Omega_\theta^2}\rho^2 f \right]_{\rho=\rho_t} <0.
\label{rangeSQ}
\end{equation}
The solution to this inequality ranges as
\begin{equation}
0< \sqrt{\lambda} f(\rho_t) < \sqrt{\omega^2- m^2-{4\Omega_\theta^2}\rho_t^2
-{n^2\over \rho_t^2}}.
\label{range}
\end{equation}
First, note that the value of the field for the physical turning point is always
smaller than the extremum value, $f(\rho_t) < f_m =  \sqrt{(\omega^2- m^2)/\lambda}$.
The field which past the maximum of $U_{\rm eff}$ never returns.
Second, in order to have a real solution for $f(\rho_t)$,
the maximum value of the argument in the square-root in Eq.~\eqref{range}
should be positive.
This gives a condition for the parameters,
\begin{equation}
\omega^2- m^2 > 4n\Omega_\theta .
\end{equation}
This is only a necessary condition for turning.
With given initial conditions at $\rho=0$,
it is not analytically tractable to see
whether or not the turning point $f(\rho_t)$
will be located in the range~\eqref{range}.
We need to search numerically for the right solution
which turns back and settles at the origin.

The numerical results for nontopological vortices
are shown in Fig.~2 for $n=1,2,3$.
The field $f$ increases initially with $f(\rho =0)=0$
and velocity determined by Eq.~\eqref{IC},
and then turns to decrease.
At large $\rho$, $f$ approaches zero with decaying velocity to zero.\footnote{In order to obtain
a long tail at large $\rho$ for numerical solutions,
a very fine-tuning for $f_1$ is required.
It is mainly due to the vulcanization term which needs tail-down, but
with an increasing coefficient $4\Omega_\theta^2\rho^2$ at large $\rho$.}
The decaying behavior is very close to the
approximate solution given in Eq.~\eqref{Whittaker}.

In Fig.~3, we plotted several types of field configurations.
Those configurations are obtained by varying the shooting parameter $f_1$
in Eq.~\eqref{IC}.

\noindent
(i) When $f_1$ is very large [Case (b)],
the field overcomes the maximum of $U_{\rm eff}$
and diverges to infinity.

\noindent
(ii) When $f_1$ is slightly larger than the nontopological one [Case (c)],
the field turns back. However, the attractive force by $\partial U_{\rm eff}$
is soon overcome by the repulsive force.
The field turns again and grows to infinity.~\footnote{The solution cannot have
more than two turning points. The proof is given in Appendix.}

\noindent
(iii) When $f_1$ is smaller than the nontopological one [Case (d)],
the field returns to the origin, but with nonvanishing velocity.
It passes the origin to the negative value, which becomes nonphysical.

\noindent
(iv) Between the configuration (c) and (d), we can always obtain the
nontopological solution by finely tuning $f_1$ in principle.

\begin{figure}[t]
\includegraphics[width=80mm]{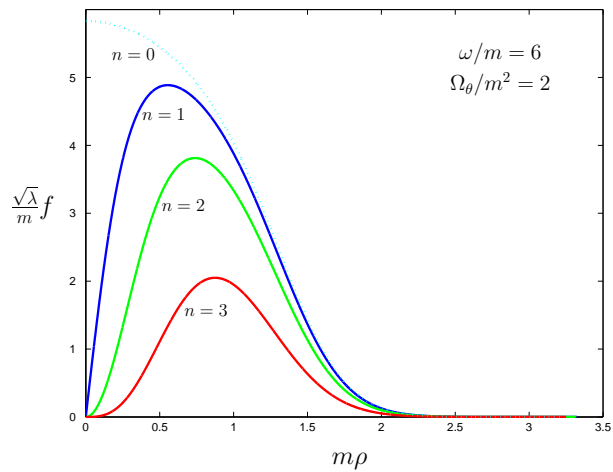}
\caption{Plot of the radial field profile
of nontopological vortices for several winding numbers,
and of Q-ball type solution ($n=0$).}
\label{fig2}
\end{figure}

\begin{figure}[t]
\includegraphics[width=80mm]{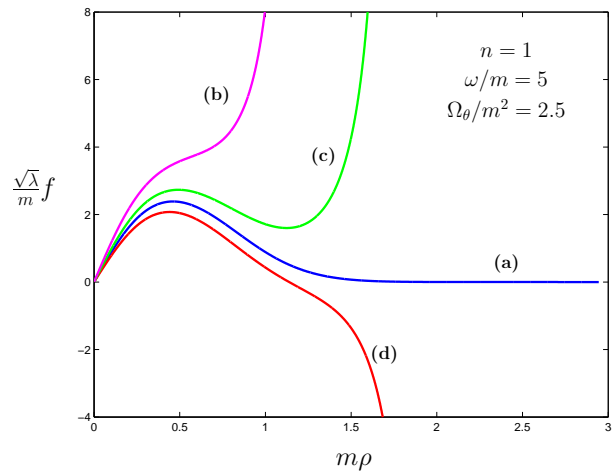}
\caption{Plot of numerical solutions for the radial profile
with different shooting parameters ($f_1$).
(a) Nontopological solution.
(b) Solution for a much larger shooting parameter than the nontopological one.
The solution monotonically increases without turning.
(c) Solution for a slightly larger shooting parameter than the nontopological one.
The solution returns before reaching the maximum of $U_{\rm eff}$.
However, the repulsive force mainly by the vulcanization term
overcomes the attractive force, so the field starts to increase to infinity.
(d) Solution for a smaller shooting parameter than the nontopological one.
The field passes the origin with nonzero velocity.
}
\label{fig3}
\end{figure}

\begin{figure}[!t]
\includegraphics[width=70mm]{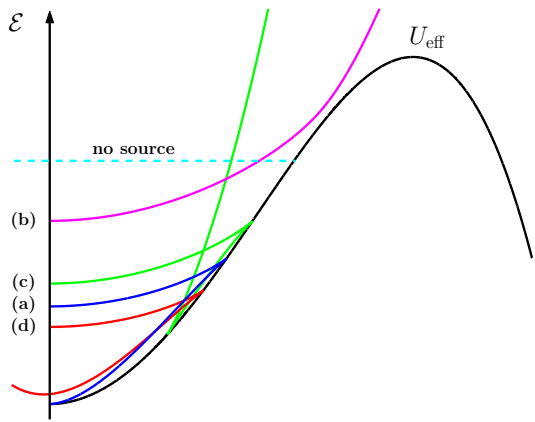}
\caption{Plot of ${\cal E}$ vs $f$ for the numerical solutions
with the same parameters in Fig.~3.
The solution (b) gains energy all the time and overcomes the potential barrier.
The solution (c) exhibits two turning points ($\cE = U_{\rm eff}$)
where the velocity becomes zero.
The solution (d) passes the origin with nonvanishing kinetic energy.
Between the solution (c) and (d), there exists a nontopological solution
for which the latter $\cE = U_{\rm eff}$ occurs at $f=0$.
}
\label{fig4}
\end{figure}

The energy consideration helps one understand the dynamical description so far.
We can define the total energy $\cE$ of the mechanical system as
\begin{equation}
{\cal E} = {1\over 2} \left( {df \over d\rho} \right)^2 +U_{\rm eff}.
\end{equation}
This energy is not a conserved quantity and changes due to driving sources.
With the aid of Eq.~\eqref{feq2}, the change-rate is given by
\begin{equation}
\frac{d\cE}{d\rho}
= -\frac{1}{\rho} \left( \frac{df}{d\rho} \right)^2
+\left(\frac{n^2}{\rho^2} + {4\Omega_\theta^2}\rho^2\right)
f \frac{df}{d\rho}.
\label{rate}
\end{equation}
The terms on the right-hand side are interpreted as sources.
If there is no source at all, the energy is conserved
and the system becomes an oscillator. (See Fig.~4.)

The first term in the right-hand side of Eq.~(\ref{rate})
always extracts energy from the system.
The second term adds energy to the system when $f$ increases,
and extracts energy when $f$ decreases.
Therefore, these two terms compete when $f$ increase,
and cooperate when $f$ decreases.
At the initial moment, the condition $f(\rho=0)=0$ implies that
the potential energy is zero (by setting $V_0=0$ in Eq.~\eqref{V}),
and that the total energy $\cE$ is solely given
by the initial velocity.\footnote{From the expression for $f$
at small $\rho$ given in Eq.~\eqref{IC},
one can see that the initial velocity is nonzero only for $n=1$
($df/d\rho (0) =f_1$; the shooting parameter becomes the initial velocity for $n=1$).
Therefore, the initial energy is zero for $n \geq 2$.}
At small $\rho$, with $f$ in Eq.~\eqref{IC},
the change-rate can be expanded as
\begin{equation}
\frac{d\cE}{d\rho} =
f_1^2~n\rho^{2n-3}
\left[ n(n-1) -{(n^2-2)(\omega^2-m^2) \over 2(n+1)} \rho^2 + {\cal O}(\rho^4)
\right],
\end{equation}
which is positive for all $n$'s.
Therefore, the total energy increases initially.
Depending on the shooting parameter $f_1$,
the competition pattern between the two terms in Eq.~\eqref{rate}
is different.

\noindent
(i) When $f_1$ is large enough, the system always gains energy and
the field $f$ overcomes the maximum of $U_{\rm eff}$ at $f_m$
owing to the acquired energy [Case (b)].

\noindent
(ii) When $f_1$ is lowered, the energy gain is not very efficient,
and the total energy $\cE$ becomes the same with $U_{\rm eff}$
before the field $f$ reaches the maximum $f_m$ [Case (c)].
The kinetic energy vanishes at that moment,
and the velocity changes its sign afterwards.
When $f$ returns to the origin,
the energy $\cE$ always decreases.
If $f_1$ is not sufficiently low,
$\cE$ hits $U_{\rm eff}$ again at $f>0$, and $f$ changes its direction.
Finally, the field evolves similarly to that in (i).

\noindent
(iii) When $f_1$ is considerably low,
the system loses energy slowly during the returning of $f$, and
the second turning point ($\cE = U_{\rm eff}$) does not arise
in the region of $f \geq 0$ [Case (d)].
The field $f$ passes the origin with nonzero negative velocity.

\noindent
(iv) Tuning $f_1$ finely between those for (c) and (d),
there exists a trajectory for which the $\cE = U_{\rm eff}$ point
occurs at $f=0$ [Case (a)].
The velocity vanishes there, $df/d\rho=0$,
and the acceleration becomes zero, $d^2f/d\rho^2 =0$,
as it was analyzed earlier.
This will occur at $\rho\to\infty$
and one gets the nontopological vortex solution.

\subsection{Q-ball type solution}
\begin{figure}[!t]
\includegraphics[width=70mm]{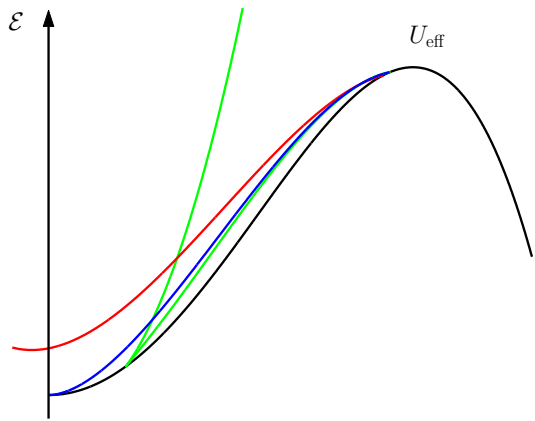}
\caption{Plot of ${\cal E}$ vs $f$ for Q-ball type solutions for
$\omega/m =5$ and $\Omega_\theta/m^2 =2.5$.
The blue line represents the solution which meets the boundary conditions
of Q-ball type.
If the shooting parameter $\xi$ is a bit smaller
(i.e., $f(0) =f_m-\xi > f_{\rm Q-ball}(0)$),
there exists a turning point and the field $f$ diverges as in Case (c) for vortices.
If the shooting parameter $\xi$ is a bit larger,
the field passes the origin as in Case (d) for vortices.
}
\label{fig5}
\end{figure}

The Q-ball is a nontopological soliton
which was first investigated in~\cite{Friedberg:1976me,TDLee}.
It possesses only the U(1) Noether charge, but no vorticity ($n=0$).
Here, we shall search a solution which satisfies the boundary conditions
of the Q-ball type.
Similarly to vortices, the radial profile of the field approaches asymptotically
the symmetry state, $f (\rho \to\infty)\to 0$, but
from a nonzero inner boundary value which is off the local extremum of $U_{\rm eff}$,
$f(\rho=0) =f_m-\xi$.
We try to search a numerical solution which satisfies these boundary conditions.
For the inner boundary,
we can perform a series expansion in the vicinity of $\rho=0$,
\begin{eqnarray}
f(\rho) &=& (f_m -\xi)
-{\lambda\over 4}\xi (f_m-\xi)(2f_m-\xi)\rho^2 \nonumber\\
&+& {\lambda\over 64}(f_m-\xi)
\left[ -16 {\Omega_\theta^2 \over \lambda} +4(\omega^2-m^2)f_m\xi
-14(\omega^2-m^2)\xi^2 +12\lambda f_m\xi^3 -3\lambda\xi^4\right]
\rho^4 +{\cal O}(\rho^6).
\end{eqnarray}
Now, the shooting parameter is $\xi$.
By varying $\xi$, we search a numerical solution which meets
the outer boundary condition.
Dynamical interpretation states that the field $f$ starting from
$f=f_m-\xi$ should settle down to the local
minimum at $\rho\to \infty$.
The numerical solution for $f(\rho)$ is plotted in Fig.~\ref{fig2},
and the energy profile is plotted in Fig.~\ref{fig5}.
The asymptotic behavior of $f$ approximates the analytic function~\eqref{Whittaker}
with setting $n=0$.
The existence of such a Q-ball type solution is very similar to
that for the nontopological vortices explained based on Fig.~\ref{fig3}.
If $\xi$ is a bit smaller than that for the Q-ball type solution,
the field $f$, which decreases initially from $f_m-\xi$, turns and
diverges to infinity. It is similar to Case (c) for vortices.
If $\xi$ is a bit larger, $f$ decreases and passes $f=0$ to negative
similarly to Case (d) for vortices.
Between them, there exists a Q-ball type solution.

For nontopological vortices and Q-ball type solitons,
the integrated energy over the cross-sectional 2D space,
\begin{equation}
E = 2\pi \int^\infty_0 d\rho \rho {\cal H},
\end{equation}
is finite, where the Hamiltonian density is given by
\begin{eqnarray}
{\cal H} = {1\over 2} \nabla_i \bar\phi \nabla^i \phi
+ V +{1\over 2}\omega^2\bar\phi\phi
+ {\Omega^2 \over 2}\tilde{x}_\mu\tilde{x}^\mu\bar\phi\phi
= {1\over 2} \left( {d f \over d\rho} \right)^2 + {n^2 \over 2} {f^2 \over \rho^2}
+ V +{\omega^2\over 2}f^2
+{2\Omega_\theta^2} \rho^2 f^2 .
\end{eqnarray}
As it was mentioned earlier, in order to have zero acceleration at the outer boundary,
$f$ should decay faster than $1/\rho^2$.
This guarantees that the integration of the above Hamiltonian density
converges.

\section{conclusions}
In this article, we investigated solitonic configurations
motivated by the vulcanized theory
which was introduced to renormalize the
$\phi_\star^4$ noncommutative field theory.
We studied the classical limit of the theory
which is distinctly different from the ordinary one.
The vulcanization term is always present in this limit,
and is described by the parameter $\Omega_\theta$
that is the ratio of the vulcanization parameter $\Omega$
and the noncommutative parameter $\theta$.
We set up a model of global U(1) symmetry
described by a complex scalar field.
Since the vulcanization term is $x$-dependent,
we considered inhomogeneous field configurations
such as vortices and Q-ball type solitons.
We performed numerical calculations
to find solutions to the field equation which satisfy
boundary conditions for those configurations.

The result shows that there can form nontopological vortices
and Q-ball type solitons, but no topological vortices.
Because of the vulcanization term,
the radial profile of the scalar field cannot
approach a nonzero constant value asymptotically,
so it cannot acquire the nonzero vacuum-expectation value
at the boundary.
However, the field can approach the nontopological boundary value
which represents the unbroken-symmetry state.
This was possible with the aid of the vulcanization term
with $\phi^4$-potential in the absence of the $\phi^6$ term
which plays the major role in the usual nontopological solitons.
The $n=0$ solution corresponds to the Q-ball type solution.
The $n \neq 0$ solutions correspond to the nontopological vortex
which has a constant angular momentum.

The integrated energy of nontopological vulcanized solitons is finite.
Since physics of the vulcanized field theory
has not been completely studied yet,
identification of the objects that resulted from the theory
with the physical objects (e.g., meson) is incomplete.
Therefore, whether or not the nontopological vulcanized solitons
are the lowest-energy objects in the theory is not yet manifest.
It is too early to discuss their stability or decay,
and needs further investigation.

The nontopological solitonic objects that we obtained here
are realized as stationary stringlike objects in 3D space in the Universe.
Our results can be applied to the stage in the early universe
where the renormalizable noncommutativity still works and its
classical limit is also viable.
We expect that the cosmological implications of our solutions
would not be much different from those
in the usual commutative theory~\cite{Lee:1986ts}.

\begin{acknowledgments}
The authors are grateful to Yoonbai Kim and Hiroaki Nakajima for useful discussions.
Y.L. was supported by the Korea Research Foundation Grant (KRF-2008-314-C00063)
funded by the Korean Government (MOEHRD).
\end{acknowledgments}

\begin{appendix}*
\section{Proof of no multi-hump solution for $f$}
\label{appen1}
Here we prove that there is no solution to~\eqref{feq2}
which has more than two local maxima in $f(\rho)$.

\noindent
{\it Proof})
Let $f(\rho)$ be a solution that has two local maxima
at $f_1= f(\rho_1)$ and $f_3= f(\rho_3)$,
and one local minimum at $f_2 = f(\rho_2)$
($\rho_1<\rho_2<\rho_3$), i.e.,
$f_2 = \min(f_1,f_2,f_3)$.
From Eq.~\eqref{rangeSQ}, the local maximum/minimum condition
can be written as
\begin{equation}
\lambda f_1^2<R(\rho_1),\quad
\lambda f_3^2<R(\rho_3),\quad
\text{and }~\lambda f_2^2>R(\rho_2),
\label{maxmin}
\end{equation}
where $R(\rho)\equiv \omega^2- m^2-4\Omega_\theta^2 \rho^2
-n^2/\rho^2$,
and we consider only $f(\rho)> 0$.
Since $R(\rho)$ is convex in $\rho$,
$R(\rho_2)$ is always larger than at least
one of the others,
$R(\rho_2)> R(\rho_1) \text{ or/and } R(\rho_3)$.
Therefore, from~\eqref{maxmin})
\begin{equation}
\lambda f_2^2 > R(\rho_2)> R(\rho_1) \text{ or/and } R(\rho_3)>
\lambda f_1^2 \text{ or } \lambda f_3^2.
\end{equation}
However, this inequality is against the initial assumption
$f_2 = \min(f_1,f_2,f_3)$ which tells that $f_2$ is the local minimum.

\end{appendix}

\end{document}